\documentstyle[a4p,12pt]{article} 

\def\mathrm{\rm} 

\newcommand{\cbar}{\mbox{${\overline c}$}} 
 
\newcommand{\bbar}{\mbox{${\overline b}$}} 
\newcommand{\bbbar}{\mbox{$b\bbar$}} 
\newcommand{\ccbar}{\mbox{$c\cbar$}}

\newcommand{\swsq}{\mbox{${\sin\theta_W}$}} 
\newcommand{\micro}{\mbox{${\mu}$}} 
\newcommand {\bea} {\begin{eqnarray}} 
\newcommand {\eea} {\end{eqnarray}} 
\newcommand {\beq} {\begin{equation}} 
\newcommand {\eeq} {\end{equation}} 
\input epsf 
\begin{document} 
\begin{titlepage}
  \begin{flushright}
  SLAC-PUB-7709\\           
  December 1, 1997
  \end{flushright}
 
\begin{center}
 {\Large {\bf Measurements of Hadronic Asymmetries in  
$e^+e^-$ Collisions at LEP and SLD}
\footnote{Presented  at the  ``International Europhysics Conference on High
                                             Energy Physics'', 
                                           19-26 August 1997, 
                                            Jerusalem, ISRAEL. }}

\vspace{1cm} 
{\large {\bf Erez\,Etzion} 
(Erez@lep1.tau.ac.il)\footnote{Work supported by Department of Energy contracts
DE-AC02-76ER00881 (Wisconsin) and DE-AC03-76SF00515 (SLAC).}}

{\it School of Physics University of Wisconsin, Madison, WI 53706, U.S.A} 

{\it and Stanford Linear Accelerator Center, Stanford University, CA94309, U.S.A}
 
currently: {\it School of Physics and Astronomy, Tel Aviv University, ISRAEL} 
 
\end{center}
\bigskip  
\vspace{2cm} 
\begin{abstract} 
High precision experimental new electroweak results  
measured at the four LEP experiments and the SLD collaboration  
are discussed.  
Heavy quark ($\bbbar$ and $\ccbar$)  
forward-backward asymmetries  
measured at LEP are presented along with polarized 
forward-backward and left-right asymmetries measured at SLD.  
The results are compared, and the combined averages are used to  
evaluate the Standard Model parameters. 
\end{abstract}

\vfill
\end{titlepage}
\clearpage

\section*{Introduction} 
 Around the $Z^0$ peak the fermion-pair are produced mainly through 
the $Z^0$ channel, where the $\gamma$ exchange contribution is very  
small. Asymmetry measurements, Forward-Backward (FB) and polarized 
asymmetries are sensitive to the right handed $Zf\bar{f}$ couplings  
complementary  to the partial widths measurements which are more 
sensitive to the left handed couplings. 
\noindent 
For unpolarized beams (LEP) the FB asymmetry, 
$A^f_{FB} = \frac{3}{4}A_eA_f$, 
is sensitive to the 
(initial) electron and the outgoing fermion couplings 
to the $Z^0$. 
 
\noindent 
Given the longitudinal polarization of the electron beams at SLD, 
one can use that knowledge to simply measure the difference 
between left and right handed cross-section,  
$A_{LR}=\frac{\sigma_L-\sigma_R}{\sigma_L+\sigma_R}=P_eA_e$, 
where 
$P_e$ is the polarization of the incident $e^-$ beam. 
One can also measure the FB polarized asymmetry, 
\bea 
\label{eq:afbpol} 
A^{pol(f)}_{FB}= 
\frac{(\sigma_{L,F}-\sigma_{R,F})-(\sigma_{L,B}-\sigma_{R,B})} 
     {(\sigma_{L,F}-\sigma_{R,F})+(\sigma_{L,B}-\sigma_{R,B})}= 
\frac{3}{4}P_eA_f. \nonumber 
\eea
 
\noindent 
While the Asymmetries expected from neutrinos, charged leptons,  
u-type quarks and d-type quarks are: 1, 0.15, 0.67 and 0.94  
respectively, 
the sensitivity of these to the weak mixing angle, $\frac{\delta A_f} 
{\delta \swsq}$ are 0, -7.9,   -3.5 and -0.6.  
For comparison all the LEP and SLD asymmetries are given in terms 
of the effective mixing angle which is defined as: 
$\sin^2\theta^{eff}_W \equiv 0.25(1-v_e/a_e)$, 
where the $v_e/a_e$ is extracted from the asymmetry measurements. 
 
\noindent
Complementary to the leptonic asymmetries which are sensitive 
to the oblique radiative corrections, 
the heavy flavour 
measurements ($R_{b/c}, A_{b/c}$) test the Standard Model (SM)~\cite{SM} 
through vertex corrections. 
The $c$ and the $b$ quarks being the heaviest quarks  
accessible at the $Z^0$ peak, may provide a potential window  
to physics beyond the SM.  
 
\noindent 
The measurements are based on about 4.4M events collected at each 
of the LEP experiments and the 200K events accumulated at SLD, produced with 
highly polarized electron beam. 
All the experiments have a good leptonic identification,  
DELPHI and SLD use Cerenkov ring  
imagine devices for particle identification, 
where the other 3 experiments use 
dE/dx  for that purpose. The $b$ and $c$ event tag is based mainly 
on the experiments vertexing capability. All LEP experiments have installed 
double sided silicon micro-strip vertex detector, providing a typical high  
momentum track impact parameter resolution of $\sim 25\micro$m,  
where SLD uses silicon pixel 
detector which its upgraded version, installed before the 1996 run, gives 
a high momentum track impact parameter resolution of $\sim 13\micro$m. 
 
\section*{Polarized Asymmetries} 
 
SLD has a new preliminary measurement of $A_{LR}$  
based on the data collected in 1996. 
The event sample, mostly consists of hadronic $Z^0$ decays, 
has 28,713 and 22,662 left- and right-handed electrons respectively. 
The resulting measured asymmetry is thus 
$A_m = (N_L - N_R) / (N_L + N_R) = 0.1178 \pm 0.0044$(stat). 
To obtain the left-right (LR) cross-section asymmetry at the SLC 
center-of-mass energy of 91.26 GeV, a very small correction 
$\delta = (0.240 \pm 0.055)\%$(syst) 
is applied which takes into account residual contamination in the 
event sample and slight beam asymmetries. 
As a result, 
\bea 
  A_{LR}(91.26~\mbox{GeV}) = 
       \frac{A_m}{\langle P_e \rangle} \left(1 + \delta\right) 
       = 0.1541   
         \pm 0.0057 \mbox{(stat)} \pm 0.0016 \mbox{(syst)}  \nonumber 
\eea 
where the systematic uncertainty is dominated by the systematic 
understanding of the beam polarization. 
Finally, this result is corrected for initial and final state radiation 
as well as for scaling the result to the $Z^0$ pole energy: 
\bea 
  A^0_{LR} & = & 0.1570 \pm 0.0057 \mbox{(stat)} \pm 0.0017 \mbox{(syst)} \nonumber 
 \\ \nonumber 
  \sin^2\theta_W^{eff} & = & 0.23025 \pm 0.00073 \mbox{(stat)} 
       \pm 0.00021 \mbox{(syst)}.   
\eea 
The 1996 measurement combined with the previous  
measurements~\cite{alr95} 
 yield: 
\bea 
\label{Equ_ALR} 
  A^0_{LR}  =  0.1550 \pm 0.0034 \;\; ; \;\;\;  
  \sin^2\theta_W^{eff}  =  0.23051 \pm 0.00043, \nonumber 
\eea 
which is  the single most precise determination of 
weak mixing angle. 
 
\noindent 
SLD has presented a direct measurement of the $Z^0$-lepton coupling 
asymmetry parameters based on a sample of 12K leptonic $Z^0$ decays 
collected  in 1993-95~\cite{afbpol}.  
The couplings are extracted from the measurement 
of the double asymmetry formed by taking the difference in number of 
forward and backward events for left and right beam polarization data  
samples 
 for each lepton species. 
This measurement has a statistical advantage of  
$(P_e/A_e)^2\sim 25$ on the LEP FB asymmetry measurements. It 
is  independent of the SLD $A_{LR}$ using $Z^0$ decays to hadrons, and 
it is the only measurement which determines $A_{\mu}$ not 
coupled to $A_e$. 
The results are: $A_e=0.152\pm 0.012(stat)\pm 0.001(sys)$,  
$A_{\mu}=0.102\pm 0.034\pm 0.002$, and $A_{\tau}=0.195\pm 0.034\pm 
0.003$ or assuming universality $A_{\ell}=0.151\pm0.011$.  
A new preliminary measurement based on the $\mu$ pairs collected at the 
1996 run $A_{\mu}(1996)=0.164\pm 0.046$ is given, and combined with  
the 93-95 yield $A_{\mu}=0.123\pm 0.027$. 
 
\noindent 
The SLD  preliminary weak mixing angle value combining  
$A_{LR}$, $Q_{LR}$ and 
leptons asymmetries measurements is: 
  $\sin^2\theta_W^{eff} = 0.23055 \pm 0.00041$, 
which is more than 3$\sigma$ below the LEP average. 
 
\section*{Heavy Quark Asymmetries} 
The heavy quark final state asymmetries at the $Z^0$ peak are  
large and more sensitive to the 
EW parameters than the leptonic asymmetries. 
Since the LEP FB asymmetries are sensitive to both the electron and the 
out going fermion coupling to the $Z^0$  
it can be interpreted in two ways: 
\begin{itemize} 
\item [1.] Assume universal SM coupling of the out going quarks and derive 
    $\swsq$, with a much higher sensitivity than the leptonic measurements  
    gaining from the high asymmetry values of $A_c \sim 0.67$ and 
    even better $A_b \sim 0.94$. 
\item [2.] Use $A_e$ value as given by  the leptonic (LEP+SLD)  
    measurements to determine the parity violation parameters, $A_{b,c}$. 
\end{itemize} 
The LEP FB asymmetries~\cite{afb1}-~\cite{afb4} 
 were measured on the $Z^0$ peak and above and  
below the peak. Off peak measurements were corrected to the peak energy 
before combining to a pole asymmetries $A^{0,f}_{FB}$. 
 
\noindent 
The SLD LR FB asymmetries~\cite{sldabc} are independent of the electron  
coupling 
and therefore provide a direct measurement of $A_{b,c}$ again with that  
statistical advantage of $(P_e/A_e)^2 \sim 25$ compared to the  
LEP measurements.  
 
\noindent 
Three different principal methods to select a sample of $\bbbar$ and  
measure the $b$ asymmetry have been presented: 
\begin{itemize} 
\item Use the charge of the e or $\mu$ in $b$ leptonic decays to sign 
     the $b$ quark direction, where the analyzing power is estimated from 
     the momentum and transverse momentum distribution studied with 
     simulation. (Used by ALEPH, DELPHI, L3, OPAL and SLD). 
\item Selection of $b$ sample based on lifetime/mass distribution, 
    and momentum weighted track charge to sign the $b$ quark direction. 
    This is a self calibrated technique where the analyzing power 
    is derived directly from the data. (Done by DELPHI, L3, OPAL and SLD). 
\item A unique method introduced by SLD using mass tag to select 
    the $\bbbar$ events and the Cerenkov Ring Imagine Detector (CRID) to  
    identify kaons. 
    The $b$ quark charge is assigned on the basis of the charge of the kaons 
    from the $B$ decay chain. 
\end{itemize} 
The new $b$ asymmetry analyses presented at 1997 are jet charge  
measurements by L3 (new), 
OPAL (final numbers - winter 97) and SLD (updated at winter 97),  
lepton tag by L3 and SLD (both updated for winter 97) and the SLD k-tag 
analysis updated for summer 97. The ALEPH preliminary jet charge analysis  
introduced 
in summer 96 which gave the single most precise determination of 
$A_{FB}^b$ has been withdrawn before the conference and a correction 
is expected. 
The LEP measurements of $A_{FB}^b$ and $A_{FB}^c$ extrapolated to the $Z^0$  
peak are given in Fig.~1 (a) and (c). The LEP measurements  
translated to a $b$ or $c$ quark asymmetry using the LEP/SLD combined  
value for $A_e=0.1505\pm0.0023$, are given in Fig.~1 (b)  
and (d) along with the SLD direct parity violating parameters $A_{b/c}$  
measurements. 
 
\noindent 
The measurement of the charm hadrons asymmetry is performed in a similar way, 
and again three principal methods were used for these analyses: 
\begin{itemize} 
\item Lepton tag similar to the $b$ asymmetry analysis used by all five 
experiments. (SLD's number updated at the winter conferences). 
\item $D^*$, $D^0$ and $D^+$ exclusive or semi-exclusive reconstruction  
     used by ALEPH, DELPHI, OPAL and SLD to select a sample enriched with  
      $c$ hadron events, and  to determine the direction of the $c$ quark  
     direction. 
\item In a new method introduced by SLD,  
       charm quarks were tagged by an inclusive mass tag based on  
       topological vertexing, resulting in efficiency of 20\% with 
       a purity of 75\%. Again the CRID was used to identify kaons and 
       the charge separation is obtained from a  
       combination of the vertex charge and sign of kaon from the $D$ 
       decay chain. 
   \end{itemize} 
\begin{figure}[p] 
\label{fig:afbb} 
\begin{minipage}{\textwidth} 
 \begin{minipage}{0.49\textwidth} 
 \epsfxsize=\textwidth 
 \epsffile{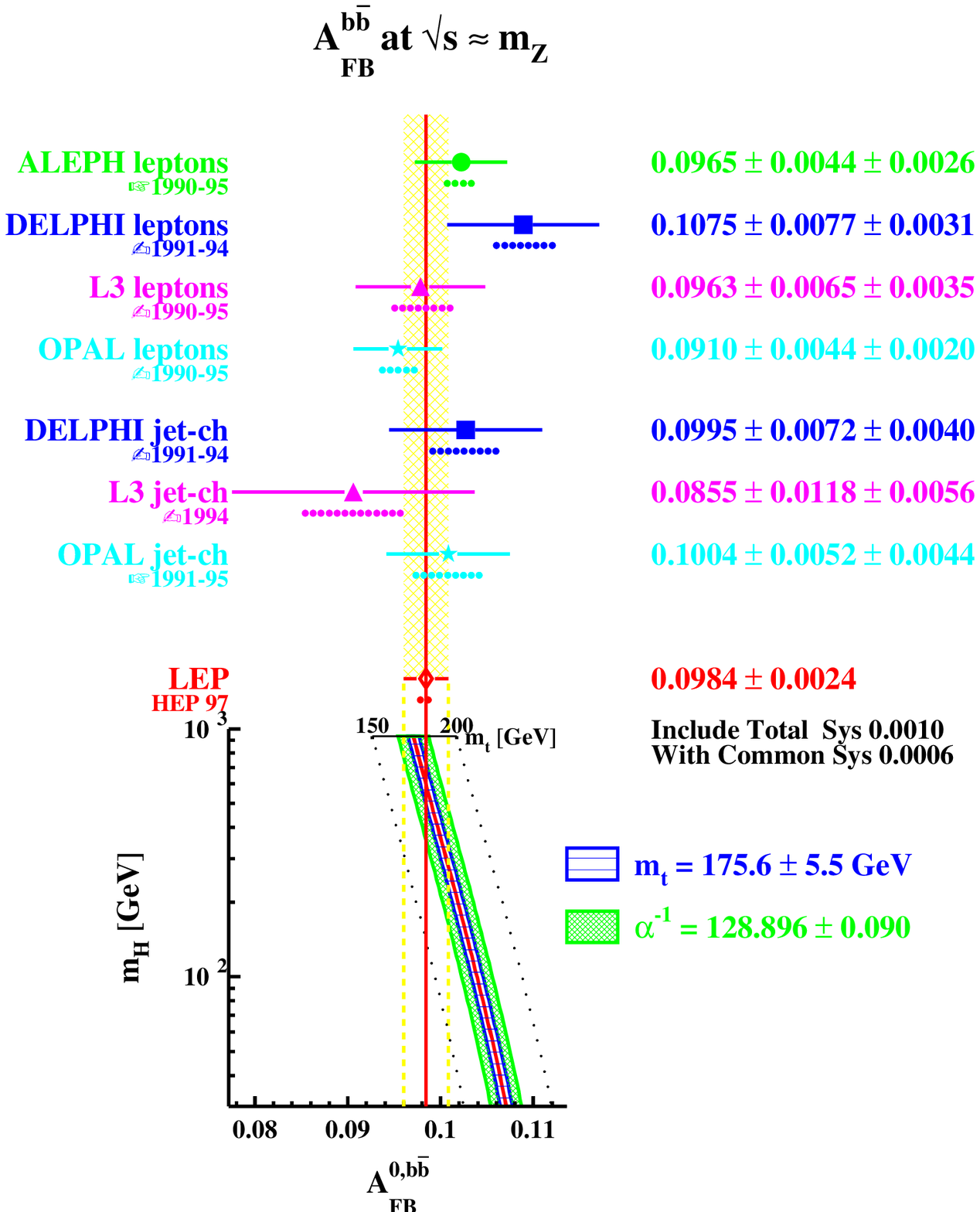} 
 \end{minipage} 
 \begin{minipage}[l]{0.49\textwidth} 
 \epsfxsize=\textwidth 
 \epsffile{ab_summer97.epsi} 
 \end{minipage} 
\end{minipage} 
 \begin{minipage}{0.49\textwidth} 
 \epsfxsize=\textwidth 
 \epsffile{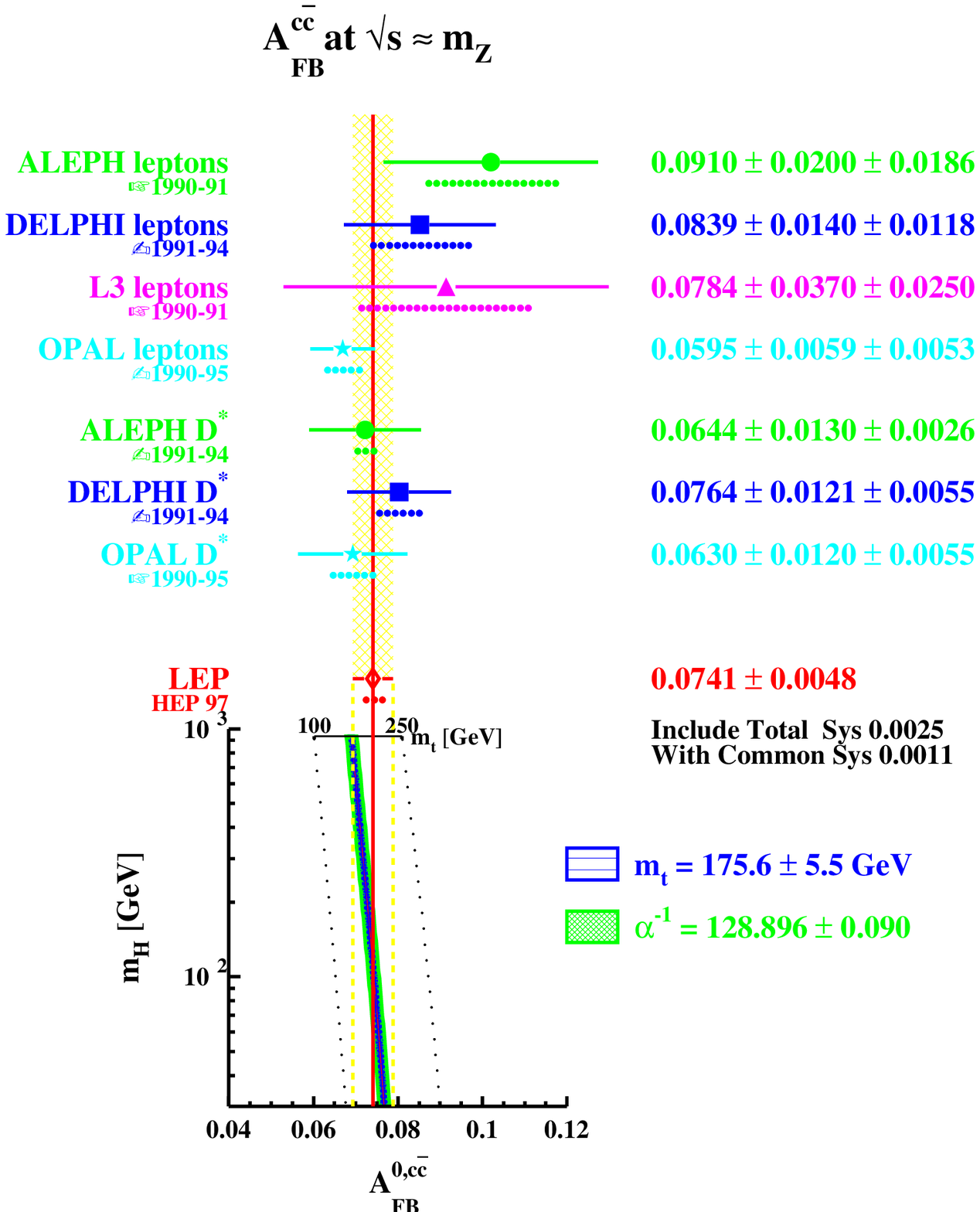} 
 \end{minipage} 
 \begin{minipage}[l]{0.49\textwidth} 
 \epsfxsize=\textwidth 
 \epsffile{ac_summer97.epsi} 
 \end{minipage} 
\caption{On the left the LEP measurements of $A_{FB}^b$ (top) and $A_{FB}^c$ 
(bottom), on the right $A_b$ and $A_c$ asymmetries as measured by SLD and LEP.} 
\end{figure} 
 
\noindent 
After calculating the overall averages, the following corrections 
are applied (using ZFITTER~\cite{ZFITTER}) in order to derive the quark pole asymmetries  
$A^{0,b}_{FB}$ ($A^{0,c}_{FB}$): 
-0.0013 (-0.0034) for the energy shift from 91.26~GeV/c$^2$ to $M_z$, 
+0.0041 (+0.0104) QED correction and -0.0003 (-0.0008) for 
$\gamma$ exchange and $\gamma$Z interference. 
QCD corrections (see e.g.~\cite{stav}) depend strongly on the  
experimental analyses. Therefore the numbers quoted for the experiments 
were already corrected for QCD effects as described in~\cite{qcd}.

\noindent 
A fit to the LEP and SLD data~\cite{ref:LEPEW-sum} gives the following combined results for 
 the electroweak parameters:
 
\begin{center}
\begin{tabular}{lll}
$R_b^0 =  0.2170\pm0.0009$        & & $R_c^0 =  0.1734\pm0.0048$  \\ 
$A^{0,b}_{FB} =  0.0984\pm0.0024$ & & $A^{0,c}_{FB} =  0.0741\pm0.0048$ \\ 
$A_b =  0.900\pm0.050$            & & $A_c =  0.650\pm0.058$ 
\end{tabular}
\end{center}
with $\chi^2/d.o.f=0.65$. The $R_b$-$R_c$ correlation is -20\%,  
the $A^{0,b}_{FB}$-$A^{0,c}_{FB}$ correlation is 13\% 
and other elements in the correlation matrix are between 1 to 8\%, 
where the parameters $A_b$ and $A_c$ have been treated as independent 
of the FB asymmetries.

\section*{The Hadronic Charge Asymmetry ${\langle Q_{FB} \rangle}$} 
\label{QFB} 
The LEP experiments~\cite{QFB1}-~\cite{QFB4}  have provided  
measurements of the average charge flow  
in the inclusive samples of hadronic decays which is  related to FB of the 
individual quarks asymmetry as following: 
\bea 
<Q_{FB}>=\sum_{quark\;\;flavour}{\delta_fA_{FB}^f 
          \frac{\Gamma_f}{\Gamma_{had}}}. \nonumber 
\eea 
The charge separation, $\delta_f$, is the average charge difference between 
quark and antiquark in an  event. The $b$ and $c$ are extracted  
from  the data, the $\delta_b$ as a by-product of the $b$ asymmetry  
measurement (self calibration)  where the charm separation is obtained  
using the  hemisphere opposite to a fast $D^{*\pm}$. Light quark  
separations are derived from MC hadronization models which is the main  
systematic source. 
The results expressed in terms of the weak mixing angle are: 
\bea 
0.2322 \pm 0.0008(stat) \pm 0.0007(sys.\;exp) \pm 0.0008 (sep.) 
\nonumber \\ 
0.2311 \pm 0.0010(stat) \pm 0.0010(sys.\;exp) \pm 0.0010 (sep.) 
\nonumber \\ 
0.2336 \pm 0.0013(stat) \pm 0.0014(sys.\;exp) \;\; (new - winter\;\; 97) 
\nonumber \\ 
0.2321 \pm 0.0017(stat) \pm 0.0027(sys.\;exp) \pm 0.0009 (sep.) 
\nonumber  
\eea 
for ALEPH, DELPHI, L3 and OPAL respectively. 
 
\section*{Light Quark Asymmetries} 
 
A measurement of the charge separation, the average charge difference 
between the quark and the antiquark hemispheres in an event,  is required 
for inclusive measurement of the FB asymmetries of individual quarks. 
When the data is used to derive the charge separation in $b$ and $c$ 
quark asymmetry measurements, for light quark this is only obtained from 
the simulation and depend on the fragmentation model calibrated to 
the data. 
ALEPH, DELPHI and OPAL have published measurements of light quark  
asymmetries in the past. 
DELPHI has presented a new measurement of strange quark asymmetry based on 
a track by track identification of fast charged kaons, and anti $b$ tag 
algorithm to reduce $c$ and $b$ contamination, resulting in: 
$A_{FB}^s=0.114 \pm 0.019 \pm 0.005$. 
OPAL has performed a new analysis using all the 90-95 data to measure 
the branching fractions of the $Z^0$ into up and down type quarks, and the  
FB asymmetry in $d$ and $s$ quark events, using high momentum 
stable particles as a tag. The OPAL light quark asymmetry numbers are: 
$A_{FB}^{d,s}=0.068\pm0.035\pm0.011$ or a correlated result  
$A_{FB}^u=0.040\pm0.067\pm0.028$. 
 
\section*{Weak Mixing Angle Measurement} 
 
The different values of $\swsq$ as obtained from the leptons 
and quark asymmetries, tau polarization and LR asymmetries 
are given in table~1. A $\chi^2$/d.o.f of 12.6/6 is obtained 
where the largest disagreement at a level of 3~$\sigma$ 
is between the most precise measurements $A_{FB}^b$ at LEP 
and $A_{LR}$ at SLD. 
 
\par\noindent 
\parbox{180mm}{ 
 \parbox{80mm}{ 
  {\hspace{4mm}  
   \begin{center} 
   \begin{tabular}{|l|c|} 
   \hline \hline 
   Observable             &  $\swsq$ \\ \hline
   Lepton FB              &  $0.23102 \pm 0.00056$ \\ 
   $\tau$ polarization    &  $0.23228 \pm 0.00081$ \\ 
   $\tau$ FB polarization &  $0.23243 \pm 0.00093$ \\ 
   $b$ FB asymmetry       &  $0.23237 \pm 0.00043$ \\ 
   $c$ FB asymmetry       &  $0.2315 \pm 0.0011$ \\ 
   Jet charge             &  $0.2322 \pm 0.0010$ \\ \hline
   LEP average            &  $0.23199 \pm 0.00028$ \\ \hline

   Left-Right (SLD)     &  $0.23055 \pm 0.00041$ \\ \hline
   {\bf LEP+SLD }     &  {\bf $0.23152 \pm 0.00023$} \\ 
   \hline \hline 
   \end{tabular} 

   {Table 1: Summary of the LEP and SLD $\swsq$ measurements.}  
  \end{center} 
  } 
 } 
 \parbox{80mm}{ 
  \parbox{80mm}{ 
   \epsfxsize 78mm  
   \epsfysize 78mm 
   \epsfbox{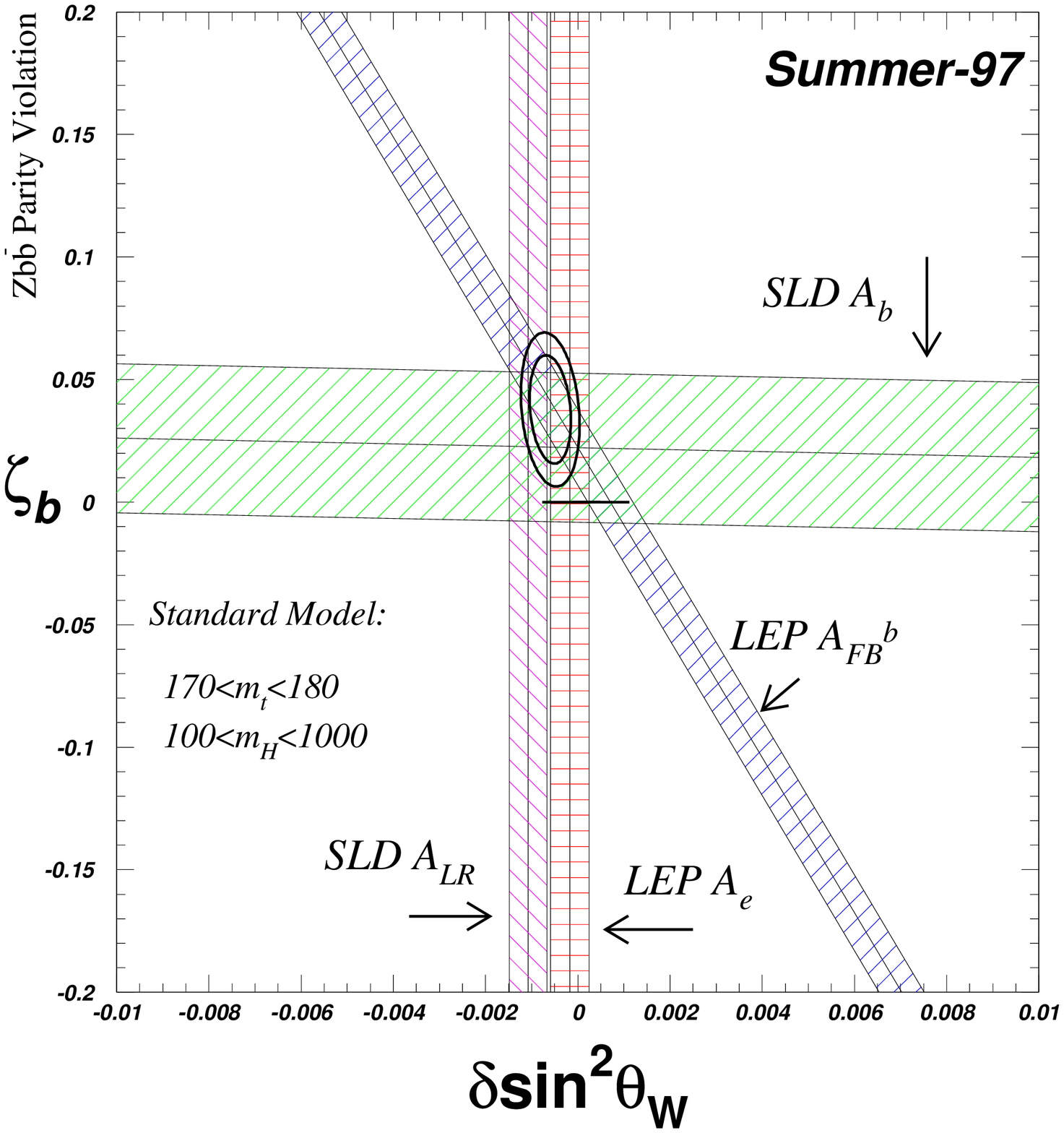} 
  } 
  \parbox{80mm}{ 
    \begin{center} 
   \parbox{80mm}{ 
     {  
      Fig. 2: The measured values of 
      $\swsq$ as derived from SLD's $A_{LR}$ and leptons asymmetries at LEP, 
   LEP $A^b_{FB}$ and SLD $A_b$ measurements plotted on $\delta \swsq$ versus 
     $Z_{\bbbar}$ parity violation plot (following Takeuchi, Grant and  
     Rosner~\cite{ref:TGR}). 
    } 
  } 
    \end{center} 
}}}  

\vspace{-5mm} 
\section*{Summary} 
   Quark final state asymmetries provide experimmental 
   powerful tools to study the SM. New measurements and some new 
   techniques were preseted at the conference. The combination of  
   many precise elctroweak results agree well with the theory predictions. 
   The current LEP+SLD average value of $A_b$ is  3~$\sigma$ lower  
   than predicted by the SM, or deriving the weak mixing angle from the  
   LEP $A_{FB}$ measurements makes the two most precise $\swsq$  
   measurements ( $A_{LR}$ and $A^b_{FB}$) disagree at a level of  
   3~$\sigma$. 
   An interesting illustration of the current LEP and SLD 
   $\swsq$ measurements plotted against the $b$ parity asymmetry 
   measurements is given in Fig.~2. 
   The use of new tagging techniques which have been introduced in  
   the $R_b$ measurements~\cite{rbeps97} 
   and analyzing the data which still has not  
   being analyzed can further improve the precision of the $b$ and  
   $c$ asymmetry measurements. SLD has started a new run and is 
   expected to reach a precision on $\delta \swsq \sim 0.0002$. 
 
\section*{Acknowledgment} 
I would like to thank the LEP and SLD collaborators and our  
colleagues from the  
LEP and SLC accelerators who brought the field to such an impressive  
level of precision. Many of the averages for this talk were derived  
by the detailed careful work of the LEP/SLD EW group~\cite{ref:LEPEW-sum}.  
It is a pleasure to thank the organizers for the interesting and successful 
conference.

%
 
\end{document}